\documentclass[12pt,preprint]{emulateapj}

\usepackage{enumerate}

\usepackage{amsmath}
\usepackage{amssymb}

\usepackage{lineno}

\usepackage{rotating}

\usepackage{soul}
\usepackage[usenames,dvipsnames]{color}

\usepackage{url}

\usepackage[multiple]{footmisc}

\usepackage[framemethod=tikz]{mdframed}

\usepackage{hyperref}
\hypersetup{
colorlinks=true,
breaklinks=true,
urlcolor= RedViolet,
citecolor=blue,
linkcolor=blue
}

\definecolor{vertdc1}{RGB}{20,89,33}
\definecolor{blood}{RGB}{193,41,41}
\definecolor{viol}{RGB}{109,10,186}
\definecolor{dgreen}{RGB}{9,95,29}
\definecolor{dorange}{RGB}{197,69,6}
\definecolor{dblue}{RGB}{43,43,179}
\definecolor{brik}{RGB}{119,15,15}

\definecolor{CNRSBlue}{RGB}{26,48,81}
\definecolor{CNRSLightBlue}{RGB}{10,141,167}
\definecolor{DarkRed}{RGB}{212,0,0}
\definecolor{DarkRed2}{RGB}{150,0,0}
\definecolor{Violet}{RGB}{122,47,214}
\definecolor{DarkViolet}{RGB}{117,36,180}

\slugcomment{}

\shorttitle{Venus low atmosphere chemical gradient}

\shortauthors{Cordier et \textit{al.}}

\begin{document}


\title{The Physical Origin of the Venus Low Atmosphere Chemical Gradient}


\author{
Daniel~Cordier\altaffilmark{1},
David~A.~Bonhommeau\altaffilmark{1},
Sara~Port\altaffilmark{2},
Vincent~Chevrier\altaffilmark{2},
S\'{e}bastien~Lebonnois\altaffilmark{3}, and 
Fernando Garc\'{i}a-S\'{a}nchez\altaffilmark{4}
}
           
\altaffiltext{1}{Universit\'{e} de Reims Champagne Ardenne, CNRS, GSMA UMR 7331,       
                 51097 Reims, France}
                              
\email{daniel.cordier@univ-reims.fr}

\altaffiltext{2}{Arkansas Center for Space and Planetary Sciences, 
                 University of Arkansas, Fayetteville, AR 72701, USA}
                 
\altaffiltext{3}{Laboratoire de M\'{e}t\'{e}orologie Dynamique (LMD/IPSL), 
                 Sorbonne Universit\'{e}s, UPMC Univ. 
                 Paris 06, ENS, PSL Research University, 
                 Ecole Polytechnique, Universit\'{e} Paris Saclay, 
                 CNRS, 75252 Paris, France}

\altaffiltext{4}{Engineering Management of Additional Recovery,
                 Mexican Petroleum Institute.
                 Eje Central L\'{a}zaro C\'{a}rdenas Norte 152, 
                 07730 Mexico City, Mexico.}
                 
\begin{abstract}
  {\noindent Venus shares many similarities with the Earth, but concomitantly, some of its
  features are extremely original. This is especially true for its atmosphere, where high pressures and temperatures are 
  found at the ground level. In these conditions, carbon dioxide, the main component of Venus' atmosphere, is a
  supercritical fluid. The analysis of {\it VeGa-2} probe data has revealed the high instability of the region located in
  the last few kilometers above the ground level. Recent works have suggested an explanation based on the existence of a vertical  
  gradient of molecular nitrogen abundances, around $5$ ppm per meter.}
  %
  {Our goal was then to identify which physical processes could lead to the establishment of this
   intriguing nitrogen gradient, in the deep atmosphere of Venus. }
  %
  {Using an appropriate equation of state for the binary mixture CO$_2$--N$_2$ under supercritical conditions,
  and also molecular dynamics simulations, we have investigated the separation processes of N$_2$ and CO$_2$ in the Venusian context.}
  %
  {Our results show that molecular diffusion is strongly inefficient, and potential phase separation is an unlikely
  mechanism. We have compared the quantity of CO$_2$ required to form the proposed gradient with what could be released by a diffuse
  degassing from a low volcanic activity. The needed fluxes of CO$_2$ are not so different from what can be measured over some terrestrial 
  volcanic systems, suggesting a similar effect at work on Venus.}\\
\end{abstract}

\noindent

\section{\label{intro}Introduction}

 Venus, sometimes considered as the ``sister planet'' of Earth, is actually very different than what titles implies. Essentially, Venus and the Earth share 
similar masses, densities, and heliocentric distances \citep{malcuit_2015}; 
other features differ significantly, and this is particularly the case for their atmospheres. For instance, at the ground level, the air of Venus
is hellish: the temperature is close to $740$ K and the pressure is in the vicinity of $90$ bars.
Because carbon dioxide dominates the atmospheric composition, with a mole fraction around $97$\%, at low altitude, the Venusian air
is a supercritical fluid. The second most abundant atmospheric compound is molecular nitrogen, with a mole fraction
of about $3$\%. Therefore, nitrogen is also in a supercritical state. 
In the phase diagram displayed in Fig.~\ref{PTprofil} we have plotted carbon dioxide and nitrogen's critical points as well as the 
pressure ($P$)--temperature ($T$) atmospheric profile measured during the descent of the {\it VeGa-2} probe \citep{lorenz_etal_2018}. 
Venus is not the only celestial body harboring supercritical fluids.
Giant planets Jupiter and Saturn, together with brown dwarfs and some exoplanets, have regions where molecular hydrogen is in the supercritical domain 
\citep{trachenko_etal_2014}. In the case of terrestrial planets, supercritical fluids could be present, but less abundant. For example, on the 
Earth, supercritical water has been found in some hydrothermal vents \citep{koschinsky_etal_2008}, and its presence was also
known for a long time in geothermal reservoirs \citep{agostinetti_etal_2017}. Additionally, high density and high temperature lead to
efficient dissolving and extracting abilities of supercritical fluids, allowing many industrial applications \citep{bolmatov_etal_2013}.\\
    Recently, the existence of a gradient of chemical composition in the Venus deep atmosphere, {\it i.e.} for layers below the altitude of $\sim 7000$ m,
has been suggested \citep{lebonnois_schubert_2017}. The abundance of nitrogen seems to decrease from $\sim 3.5$\% at $7000$\~m, to zero
at ground level, yielding an average gradient of about $5$ ppm m$^{-1}$. If accepted as real, the proposed gradient may have two origins: 
(1) a N$_2$--CO$_2$ separation due to specific fluid features (to be identified, these can be categorized as ``intrinsic'' origins) and
(2) phenomena related to Venus surface properties, such as the continuous release of some amount of carbon dioxide from the Venusian crust 
or molecular nitrogen entrapping, which can be referred as ``extrinsic origins.''
As a very first attempt, we have checked that the apparent composition gradient could not simply be an effect of the  compressibility 
of the CO$_2$--N$_2$ mixture. For this purpose, we have employed an equation of state \citep[EoS;][]{duan_etal_1996}
specifically developed for mixtures like CO$_2$--N$_2$, under supercritical conditions.
The compressibility factor \citep[see for instance][]{zucker_biblarz_2002}
$Z$ measures the deviation of real gas compressibility from an ideal 
behavior, and a value around unity shows 
a compressibility similar to that of an ideal gas. We found that the compressibility factor $Z$ stays around $1$ in the deep atmosphere of Venus.
This result, also supported by laboratory measurements \citep{mohagheghian_etal_2015}, has already 
been found by \cite{lebonnois_schubert_2017}, and we confirm it.\\
    Although the fluid under consideration here seems to exhibit typical compressibility properties, in Sec.~\ref{separation}
we discuss the possibility of a separation of N$_2$ and CO$_2$, induced by gravity, and facilitated by some potential properties 
of our supercritical mixture. In Sec.~\ref{disconcl} we discuss the role of ``extrinsic effects'' like crust release of carbon dioxide,
before concluding on a plausible scenario.

\section{The Possible Separation of N$_2$ and CO$_2$ under Supercritical Conditions}
\label{separation}

\subsection{The Effect of Molecular Diffusion}
\label{moldiff}

   For almost a century, chemical composition variations have been known among terrestrial reservoirs of hydrocarbons, i.e.,
containing gases or petroleum \citep{sage_lacey_1939}, or both. These variations are horizontal, i.e., the composition
of the mixture changes from one well to another, but also vertical. In the latter case, for a given well, the abundances of species 
evolve with depth.
The emergence of such a gradient of composition is due to the effect of gravity and to thermodiffusion. In this context,
the convection is not very efficient because the fluid is trapped in a porous medium.
A vast collection of literature is dedicated to this topic, and informative albeit not exhaustive
reviews may be found elsewhere 
\citep{thomas_2007,obidi_2014}. Here, we focus our attention on the physics, based on molecular diffusion, of the vertical 
compositional grading. 
Our goal is to evaluate the efficiency of molecular diffusion under supercritical conditions.
If the air mixing, due to atmospheric circulation, is slow or inefficient enough, the deep atmosphere of Venus could be the subject
of analogous physical processes, leading to a gradient of nitrogen concentration. This assumption may also be regarded as an ideal
or limiting case.\\
   In a fluid medium undergoing a gradient of composition, temperature, or pressure, fluxes of matter appear. These transport 
mechanisms are ruled by the physics of irreversible physical processes. The best-known laws in this field are certainly the Fick's
law of molecular diffusion and its Fourier counterpart for thermal conduction.
%
\begin{figure}[!t]
\begin{center}
\includegraphics[width=8 cm, angle=0]{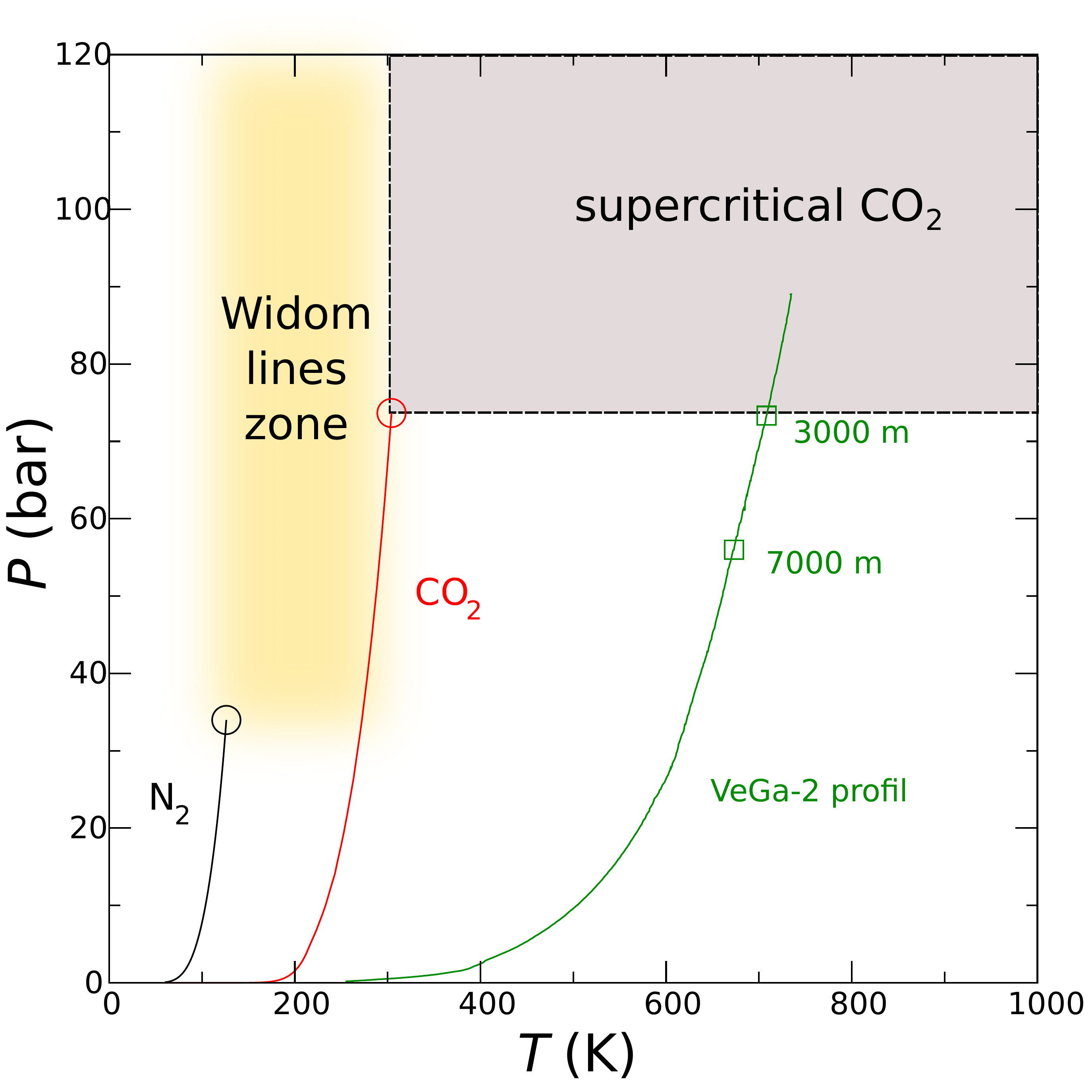}
\end{center}
\caption[]{\label{PTprofil}Simplified phase diagram of CO$_2$: the position of the critical point of pure CO$_2$ has been plotted (red circle), 
           together with the liquid-vapor equilibrium line (red line); the supercritical CO$_2$ regions are shaded gray. 
           The pressure (in bars) is denoted $P$, while $T$ represents the temperature (in kelvins) of the system.
           Similarly, 
           the N$_2$ critical point and liquid-vapor equilibrium line are shown in black.
           When some amount of molecular nitrogen is added to pure CO$_2$, the critical point of the mixture is shifted, 
           leading to a position within the yellow-shaded zone \citep{vanKonynenburgand_scott_1980,goos_etal_2011}.
           For small fractions of N$_2$, the position of the critical point is shifted by a few bars and a few Kelvin 
           \citep{goos_etal_2011}.
           The green line refers 
           to the pressure--temperature profile of Venus atmosphere derived from measurements performed by the {\it VeGa-2} probe \citep{lorenz_etal_2018}.}
\end{figure}
%
  In the general case, the total diffusion mass flux $\overrightarrow{\jmath_1}$ (kg m$^{-2}$ s$^{-1}$) of one of the two components making a binary 
mixture may be written as \citep{bird_etal_1960,ghorayeb_firoozabadi_2000}
\begin{equation}
\label{j1_gene}
\begin{split}
  \overrightarrow{\jmath_1}= -\rho & D_{12} \frac{M_1 M_2}{\bar{M}^2} \, 
         \left\{\left.\frac{\partial \mathrm{ln}\, f_1}{\partial \mathrm{ln}\, x_1}\right|_{P,T} \overrightarrow{\nabla}x_1 \right.\\
          & \left.+ \frac{x_1}{RT} \left(\bar{V}_1 - \frac{M_1}{\rho}\right) \overrightarrow{\nabla}P  
                      + \frac{k_{T \, 1,2}}{T} \, \overrightarrow{\nabla}T \right\}
\end{split}
\end{equation}
with $\rho$ being the density (kg m$^{-3}$) of the mixture, $D_{12}$ being the Fickean diffusion coefficient (m$^{2}$ s$^{-1}$) of species ($1$) in ($2$), 
$M_1$ and $M_2$ are the respective molecular weights (kg mol$^{-1}$) of the involved species, and $\bar{M}$ is the average molecular weight of the system: 
$\bar{M}= x_1 M_1 + (1-x_1) M_2$. The fugacity, which measures non-ideal effects for a real gas, is represented by $f_1$ for compound ($1$). Adopting a typical
notation, $x_1$ and $\bar{V}_1$ are the mole fraction of species ($1$) and its molar volume (m$^{3}$ mol$^{-1}$), and $P$ and $T$ are respectively the local 
pressure (Pa) and temperature (K). In the last term of Eq.~(\ref{j1_gene}),  the thermal diffusion ratio $k_{T \, 1,2}$ (dimensionless) is a function of 
$\alpha_{T\, 1,2}$, the thermal diffusion coefficient (dimensionless), according to the formula $k_{T \, 1,2}= \alpha_{T\, 1,2} x_1 (1-x_1)$.\\
  In the phenomenological Eq.~(\ref{j1_gene}), we recognize three terms that correspond to three different processes.
The first term represents the well known molecular Fickean diffusion. The second term is for pressure diffusion, which 
may lead to gravity segregation. 
The last term represents thermal diffusion (called the Soret effect for liquids), which is the tendency for species of a convection-free mixture
to separate under temperature gradient. 

     Using the pressure and temperature gradients, provided by the {\it VeGa-2} probe \citep{lorenz_etal_2018} 
under steady-state conditions ({\it i.e.} $j_1=0$) to solve Eq. (\ref{j1_gene}) allows the derivation of the chemical composition as a function of
altitude $z$. In a first approach, if we assume the Venus deep atmosphere fluid acts like an ideal gas, the equation to be
integrated is simply
\begin{equation}
\label{steady_state_ideal}
\begin{split}
  \frac{\partial x_1}{\partial z} = -\frac{x_1}{P} & \left(1-\frac{M_1}{\bar{M}}\right) \left.\frac{\partial P}{\partial z}\right|_{\rm\tiny VeGa2}
             \\&-\frac{\alpha_{T \, 1,2}}{T} \, x_1 (1-x_1) \, \left.\frac{\partial T}{\partial z}\right|_{\rm\tiny VeGa2}
\end{split}
\end{equation}
In our approach, nitrogen is our chosen compound ($1$).
We recall that, for a mixture of ideal gases, the fugacity is simply $f_1= x_1 P$, leading to 
$(\partial \, \mathrm{ln} \, f_{1}/\partial \, \mathrm{ln} \, x_1) = 1$. The partial molar volume is, for all involved species, $RT/P$; and 
the density can be written as $\rho= P\bar{M}/RT$.
     Thanks to the kinetic theory of gases \citep{chapman_cowling_1970}, the thermal diffusion coefficient $\alpha_{T \, 1,2}$ may 
be estimated for the system N$_2$--CO$_2$. In the general case, $\alpha_{T \, 1,2}$ may be positive or negative, depending on the respective 
masses of compounds of interest. For N$_2$--CO$_2$, using Chapman \& Cowling's approach, we found this coefficient to be negative.
As a consequence, heavier molecules, {\it i.e.} CO$_2$, should
gather in the coldest regions at high altitude. At the same time, the pressure term plays the opposite role, by enriching the highest layers in N$_2$.
However, our numerical simulations show that thermal diffusion remains notably smaller than pressure diffusion, with thermal flux around $30$\% of
pressure flux.\\
 By integrating Eq. (\ref{steady_state_ideal}) from the top of the Venus deep atmosphere, {\it i.e.} from an altitude of $7000$ m, down to the 
surface, we obtained a nitrogen mole fraction gradient of $\sim 0.6$ ppm m$^{-1}$. This value is roughly one order of magnitude lower than the
expected gradient \citep{lebonnois_schubert_2017} of $\sim 5$ ppm m$^{-1}$.\\
 
       Now we can turn to a more realistic model; taking into account non-ideal effects, the relevant equation is then 
\citep{ghorayeb_firoozabadi_2000}
\begin{equation}
\label{steady_state_real}
\begin{split}
\left.\frac{\partial \, \mathrm{ln} \, f_{1}}{\partial \, \mathrm{ln} \, x_1}\right|_{P,T} \frac{\partial x_1}{\partial z} & +
 \frac{x_1}{RT} \left(\bar{V}_1 - \frac{M_1}{\rho}\right) \left.\frac{\partial P}{\partial z}\right|_{\rm\tiny VeGa2}\\
 &+\frac{\alpha_{T \, 1,2}}{T} \, x_1 (1-x_1) \, \left.\frac{\partial T}{\partial z}\right|_{\rm\tiny VeGa2}
 = 0
\end{split}
\end{equation}
The pressure gradient can then be expressed as a function of density and Venus' gravity $g_V$; we then get
\begin{equation}
\begin{split}
\left.\frac{\partial \, \mathrm{ln} \, f_{1}}{\partial \, \mathrm{ln} \, x_1}\right|_{P,T} \frac{\partial x_1}{\partial z} = 
 \frac{x_1}{RT} & \left(\bar{V}_1 - \frac{M_1}{\rho}\right) \rho \, g_V \\
                & - \frac{k_{T, 12}}{T} \frac{\partial T}{\partial z}
\end{split} 
\end{equation}
  With this new equation, we can expect the enhanced composition gradient $\partial x_1/\partial z$. Indeed,
one of the criteria \citep{myerson_senol_1984} used to determine the critical point position is $\partial \mu_1/\partial x_1= 0$. 
This latter condition is essentially equivalent \citep{taylor_krishna_1993} to $\partial \mathrm{ln} f_1/\partial \mathrm{ln}\,x_1= 0$. 
Then, in the supercritical region, when the representative point of the system approaches the critical point,
this derivative tends to zero. At the same time, the quantity $\bar{V}_1 - M_1/\rho$ generally has a non-zero finite value. 
As a consequence, the gradient $\partial x_1/\partial z$ could take very large values when the system is in the vicinity of the critical point. 
Of course, this property is unchanged if the thermodiffusion term is taken into account. In order to derive quantitative estimations we have employed
the EoS \citep{duan_etal_1996} developed for the system CO$_2$--N$_2$ under supercritical conditions, and already used for our computations of
the compressibility factor $Z$. The fugacity of nitrogen $f_1$, and its derivative, are obtained from the fugacity coefficient $\Phi_1=f_1/P_1$ provided
by the EoS. Here, the quantity $P_1$ denotes the partial pressure of nitrogen: $P_1= x_1 P$. The partial molar 
volume of N$_2$ comes from the equation {\color{dblue}(see the Appendix \ref{DerivMolVol}. for derivation)}
\begin{equation}
\label{V1}
  \bar{V}_1 = \frac{R T}{P} \, \left(1 + P \, \left.\frac{\partial\mathrm{ln} \Phi_{1}}{\partial P}\right|_{T,x_{1},x_{2}} \right)  
\end{equation}
  We made comparisons, using the pressure--temperature profile acquired by {\it VeGa-2}, between quantities evaluated from
the ideal gas EoS and those calculated with the help of the more advanced Duan's EoS. Not surprisingly, we found negligible differences remaining
below a few percent. Concerning the derivative of the fugacity, we consistently obtained values around unity. Our
simulations have shown that $\partial\mathrm{ln} f_{1}/\partial\mathrm{ln} x_{1}$ tends to zero only in the close neighborhood of the binary
mixture critical point. 
   The Venusian atmosphere seems too far from this critical point to account for the effect expected for tiny  values of 
$\partial\mathrm{ln} f_{1}/\partial\mathrm{ln} x_{1}$, 
{\it i.e.} a very 
large gradient of chemical composition. In summary, our equilibrium model, taking into account non-ideal effects, leads to a gradient similar to 
the value already obtained for an ideal gas, one order of magnitude lower than what it is needed to explain the observations.\\

\begin{figure}[!t]
\begin{center}
\includegraphics[width=6 cm, angle=0]{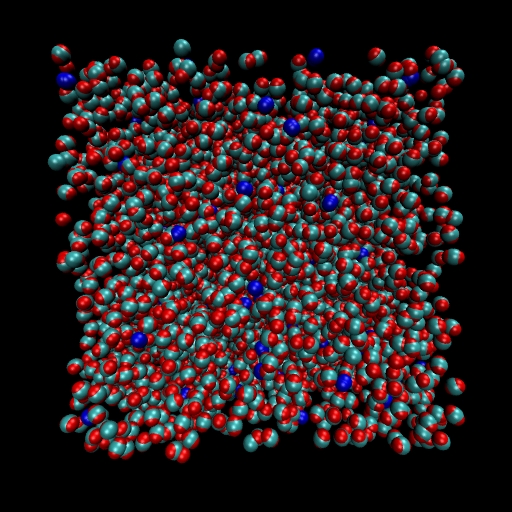}
\end{center}
\caption[]{\label{molecularview}Supercritical fluid representing Venus atmosphere (97\% CO$_2$/3\% N$_2$) at 
           $T= 740$~K with $\rho = 64$~kg/m$^3$. The simulation box is composed of $4850$ CO$_2$ molecules 
           (carbon atoms in cyan and oxygen atoms in red) and $150$ N$_2$ molecules (nitrogen atoms in blue).}
\end{figure}

  Up to this point, we have left out the question of timescales. The equilibrium states described above need a certain amount of time to be 
reached by the system. Since the transport processes considered in our model rely mainly on molecular diffusion, it is quite easy to make a 
rough estimation of the associated timescale. If $H$ is the typical size of a system 
containing a fluid, the order of magnitude of the timescale $\tau_{\rm diff}$ associated to diffusion processes is given by 
$\tau_{\rm diff} \sim H^2 / D$ where $D$ is the molecular diffusion coefficient. For the behavior of a solute, here nitrogen, 
dispersed in a volume of supercritical carbon dioxide, the value $D_{\rm N_2-CO_2}$ (m$^2$ s$^{-1}$) of this coefficient may be evaluated with 
the help of the Wilke--Chang equation \citep{wilke_chang_1955,sassiat_etal_1987} which is essentially an empirical modification of the 
Stokes--Einstein relation
\begin{equation}\label{ModifStokesEinstein}
D_{\rm N_2-CO_2} \simeq 7.4\times 10^{-15} \frac{T M_{\rm CO_2}^{1/2}}{\eta_{\rm CO_2} V_{\rm N_2, b}^{0.6}}
\end{equation}
  where $T$ is the temperature, the molecular weight of CO$_2$ is $M_s$ (44~g~mol$^{-1}$), the viscosity of supercritical carbon dioxide
is denoted $\eta_{\rm CO_2}$ (Pa~s), while the molar volume of nitrogen at its boiling temperature, under atmospheric pressure, is 
represented by $V_{\rm N_2, b}$. Here, we have $V_{\rm N_2, b} \simeq 35$~cm$^3$~mol$^{-1}$. 
The supercritical carbon dioxide viscosity $\eta_{\rm CO_2}$ may be estimated with the help of the Heidaryan {\it et al.}'s correlation 
\citep{heidaryan_etal_2011}; for $T= 700$~K and $P= 100$~bar, we got $\eta_{\rm CO_2} \sim 3 \times 10^{-5}$~Pa~s. All in all,
we obtained $D_{\rm N_2-CO_2} \sim 10^{-7}$~m$^2$~s$^{-1}$, a value that is roughly consistent with measurements for other simple
molecules, like acetone or benzene, spread in supercritical CO$_2$ \citep{sassiat_etal_1987}, for which the diffusion coefficients
show values slightly above $10^{-8}$~m$^2$~s$^{-1}$.\\
    In order to go beyond this first estimation, we have performed molecular dynamics (MD) simulations of CO$_2$--N$_2$ mixtures,
employing the open-source package GROMACS\footnote{GROningen MAchine for Chemical Simulations}$^{,}$\footnote{\url{http://www.gromacs.org}}
version 2018.2 \citep{abraham_etal_2015}. 
   The MD simulations have been carried out in the $NVT$ ensemble (\textit{i.e.} the canonical statistical ensemble) where the number of molecules $N$, 
the volume $V$ (or equivalently the density $\rho$), and the temperature $T$ of the system are constant. This ensemble has been selected for 
two main reasons: (i) the density of supercritical carbon dioxide, which should not significantly depart from that of the supercritical CO$_2$(97\%)--N$_2$(3\%) 
mixture, is tabulated in the NIST database at the pressures and temperatures representative of the Venusian atmosphere, unlike the isothermal compressibility 
that would be needed for performing MD simulations at constant pressure ($NPT$ ensemble) ; (ii) diffusion coefficients are well defined in the $NVT$ 
ensemble because the simulation box volume is not prone to fluctuations, and equilibrations in this thermodynamic ensemble are also less subject to 
numerical instabilities. The model system used for MD simulations in the $NVT$ ensemble includes a total of $5000$ molecules: $4850$ CO$_2$ molecules together 
with $150$ N$_2$ molecules (see Fig.~\ref{molecularview}), then respecting the molecular abundances of the deep Venus atmosphere. 
In particular, the conditions relevant for two altitudes have been taken into account: $740$ K (density: $64$ kg m$^{-3}$) and $673$ K (density: $44$ kg m$^{-3}$), 
corresponding respectively to the ground level and an altitude of $\sim 7000$~m. 
Molecular dynamics uses  the principles of classical mechanics to predict not only the positions and velocities of molecules as a function of time, 
but also transport properties (\textit{e.g.}, diffusion coefficients and viscosities), structural arrangements of molecules through the computation 
of radial distribution functions $g(r)$, free energies, and more generally any thermodynamic or dynamic quantity available from simulations at the 
microscopic scale. A broad variety of MD methods exist, but in the context of the Venusian atmosphere, we focused on force-fields methods where 
the potential energy of the system is a sum of intramolecular interactions (C--O and N--N bonds, O--C--O angles) and intermolecular, that is non-bonding, 
interactions (Coulomb interactions and van der Waals interactions described by Lennard--Jones potentials).
The carbon dioxide intermolecular parameters (\textit{i.e.} the C--C well depth $\epsilon_{\rm CC}$, the O--O well depth $\epsilon_{\rm OO}$, the C--C diameter 
$\sigma_{\rm CC}$, the O--O diameter $\sigma_{\rm OO}$, and the partial charges $q_{\rm C}$ and $q_{\rm O}$ on carbon and oxygen atoms), the C--O bond length 
($r_{\rm CO}^0$), and the O--C--O angle ($\theta_{\rm OCO}^0$), have been taken from the TraPPE force field \citep{potoff_robinson_2001}. 
The CO$_2$ intramolecular force constants ($k_{\rm CO}^b$ for the C--O bond and $k_{\rm OCO}^\theta$ for the O--C--O angle) have been provided by the 
CHARMM27\footnote{\url{https://www.charmm.org}}
force field \citep{bjelkmar_etal_2010}. 
%
%
For molecular nitrogen, the intermolecular parameters (\textit{i.e.} the N--N well depth $\epsilon_{\rm NN}$ and the N--N diameter $\sigma_{\rm NN}$) have also 
been taken from the TraPPE force field but the N-N bond length ($r_{\rm NN}^0$) and force constant ($k_{\rm NN}^b$) have been derived from the accurate analytic 
potential-energy curve proposed by \cite{leroy_etal_2006} for the ground electronic state of N$_2$. 
The partial charge $q_N$ has been taken equal to zero since N$_2$ is homonuclear.
The van der Waals interactions between different atoms $i$ and $j$ (\textit{e.g.}, $\epsilon_{\rm CN}$ and $\sigma_{\rm CN}$ for Lennard--Jones interactions 
between one carbon atom of CO$_2$ and one nitrogen atom of N$_2$) are built from the Lorentz--Berthelot mixing rules: 
$\sigma_{ij} = (\sigma_{ii} + \sigma_{jj})/2$ and $\epsilon_{ij} = \sqrt{\epsilon_i\epsilon_j}$.
All the force-fields parameters used in this work are gathered in Table~\ref{MDparam}. 
The ``cutoff distance'' for intermolecular interactions has been set to $4.5$~nm, a distance $3-4$ times larger than the typical $1-1.5$~nm used in 
conventional chemical applications, to ensure that atomic correlations that may extend to $\sim 1$~nm are not influenced by 
the cutoff definition.
Prior to starting MD simulations, the $5000$ molecules ($4850$ CO$_2$ and $150$ N$_2$) are randomly placed in a cubic box (edge length: 
$a= 17.8\rm\,nm$ at $T = 740\rm\,K$ and $a= 20.2\rm\,nm$ at $T=673\rm\,K$) with boundary conditions to model an infinite-sized fluid.  
At this stage the configuration of the system is not physical due to the random atomic positions; an optimization of these positions is performed 
to locate the system in the vicinity of a local potential-energy minimum, with this new configuration serving as the input configuration for the subsequent 
$NVT$ simulations. 
The equilibrium state is roughly reached after $50$ nanoseconds by using a time step of $1$ femtosecond. 
However, equilibration is carried on during $100$ additional nanoseconds for the sake of checks. 
The diffusion coefficients are derived from the MD simulations by computing the mean-squared displacements of molecules over a $1$ nanosecond production 
run, printing meaningful data every $1$ picosecond.\\
Following this approach, the carbon dioxide and molecular nitrogen diffusion 
coefficients we obtained are respectively for ground conditions
\begin{equation}
D_{\rm\scriptscriptstyle CO_2}^{\rm\scriptscriptstyle MD}(z=0)= 7.08 \pm 0.14 \times 10^{-7} \;\; \mathrm{m^2 \, s^{-1}}
\end{equation}
\begin{equation}
D_{\rm\scriptscriptstyle N_2}^{\rm\scriptscriptstyle MD}(z=0)= 8.97 \pm 0.64 \times 10^{-7} \;\; \mathrm{m^2 \, s^{-1}}
\end{equation}
while at an altitude of $\sim 7000$~m
\begin{equation}
D_{\rm\scriptscriptstyle CO_2}^{\rm\scriptscriptstyle MD}(z=7000)= 10.07 \pm 0.02 \times 10^{-7} \;\; \mathrm{m^2 \, s^{-1}}
\end{equation}
\begin{equation}
D_{\rm\scriptscriptstyle N_2}^{\rm\scriptscriptstyle MD}(z=7000)= 12.20 \pm 0.29 \times 10^{-7} \;\; \mathrm{m^2 \, s^{-1}}
\end{equation}
The indicated errors are the internal statistical uncertainties of our computations.
%
\begin{table}[!t]
\begin{center}
\caption{\label{MDparam}Our Molecular Dynamics Simulations parameters.}
\vspace*{0.5cm}
\begin{tabular}{ll}
CO$_2$                                      & TraPPE         \\ 
\hline \\
$\epsilon_{\rm\scriptstyle CC}$ (kJ mol$^{-1}$)    &  { }0.22449       \\
$\epsilon_{\rm\scriptstyle OO}$ (kJ mol$^{-1}$)    &  { }0.65684       \\
$\sigma_{\rm\scriptstyle CC}$ (nm)          &  { }0.280         \\
$\sigma_{\rm\scriptstyle OO}$ (nm)          &  { }0.305         \\
$q_{\rm\scriptstyle C}$                     &  { }0.70          \\
$q_{\rm\scriptstyle O}$                     & -0.35          \\
$r_{\rm\scriptstyle CO}^0$ (nm)             &  { }0.116         \\
$\theta_{\rm\scriptstyle OCO}^0$ ($\deg$)   & 180            \\ 
\hline
& CHARMM 27 \\ 
\hline \\
$k_{\rm\scriptstyle CO}^b\, (\rm kJ \, mol^{-1 } \, nm^{-2})$     & 784884.928 \\
$k_{\rm\scriptstyle OCO}^{\theta}\, (\rm kJ \, mol^{-1})$  & 25104     \\ 
\\[0.4cm]

N$_2$                                        & TraPPE     \\ 
\hline \\
$\epsilon_{\rm\scriptstyle NN}$ (kJ mol$^{-1}$)                     & 0.29932    \\
$\sigma_{\rm\scriptstyle NN}$ (nm)                           & 0.331      \\  
\hline
                                                             &  Additional N$_2$ parameters\\ 
\hline
$q_{\rm\scriptstyle N}$                                      & 0          \\
$r_{\rm\scriptstyle NN}^0$ (nm)                              & 0.1097679  \\
$k_{\rm\scriptstyle NN}^b$ ($\rm kJ\, mol^{-1}\,nm^{-2})$               & 1388996.32 \\ 
\hline
\end{tabular}
\end{center}
\end{table}
%
Essentially, our MD simulations lead to molecular diffusion coefficients around $10^{-6}$ m$^2$ s$^{-1}$, roughly one order of magnitude larger than
those found with our previous crude estimation based on Eq.~(\ref{ModifStokesEinstein}). These results seem to exclude any extraordinary and unexpected 
behavior where the diffusion coefficient of nitrogen would have been extremely high.\\
Then, adopting the range $10^{-7}-10^{-6}$~m$^{2}$~s$^{-1}$ and taking $H \sim 7\times 10^3$~m, for the deep Venusian atmosphere the derived
timescales are $\tau_{\rm diff} \sim 10^{13}--10^{14}$ s, corresponding to $1.6$--$16$~Myr. The Global Circulation Model of the 
atmosphere of Venus \citep{lebonnois_schubert_2017} shows a dynamical time of homogenization $\tau_{\rm dyn}$, of deep atmosphere layers, of about 
$20$ Venus days, {\it i.e.} $\tau_{\rm dyn} \sim 2 \times 10^{8}$~s. The duration of the possible separation process of
CO$_2$ and N$_{2}$ has to be much shorter than $\tau_{\rm dyn}$. 
   Clearly, in our case, even under supercritical conditions, the diffusion 
of molecular nitrogen would require too much time to form a noticeable compositional gradient.
In addition, we have previously shown that the 
gradient obtained at equilibrium cannot account for the $5$~ppm~m$^{-1}$ suggested in previous works \citep{lebonnois_schubert_2017}.
Finally, we would like to emphasize that the actual molecular diffusion coefficient $D_{\rm 12, act}$, of a species (1) in a real fluid is the 
product of $D_{12}$ (see also Eq.~\ref{j1_gene}), the molecular diffusion coefficient in the corresponding ideal gas, by the derivative of the fugacity
$\partial\mathrm{ln} f_{1}/\partial\mathrm{ln} x_{1}$. Of course, if the latter tends to zero, then $D_{\rm 12, act}$ has the same behavior.
As a consequence, the time required to get the equilibrium becomes infinite. As we can see, one more time, we have an argument against a 
N$_2$--CO$_2$ separation, based on molecular diffusion.

\subsection{Macroscopic separation mechanism of CO$_2$ and N$_2$}
\label{Macro}

   Facing the question of Venus' atmosphere chemical gradient, an alternative scenario could be the formation of CO$_2$-enriched droplets 
at some altitude, prior to their fall to the ground. Such a mechanism could easily impoverish high altitude layers in CO$_2$, whereas layers
close to the surface could be enriched in the same compound. This possibility requires a kind of ``phase transition'' that could form sufficiently
large droplets. This issue will be discussed momentarily, but we would like to examine first whether the timescales related to this
scenario could be compatible with dynamical timescales.
   On the Earth, the raindrops have a typical size around $1$ mm \citep{vollmer_mollman_2013}. For such particles of fluid, falling under
gravity, the sedimenting velocity \citep{pruppacher_klett_2010} $V_{\rm s}$ is given, if we neglect the 
small slip-correction, by the Stokes' law
\begin{equation}
    V_{\rm s}= \frac{D_1^2 g (\rho_1-\rho_2)}{18 \eta_2}
\end{equation}
with being $D_1$ the diameter (m) of the particle, $g$ being the gravity, $\rho_1$ and $ \rho_2$ being respectively the density (kg m$^{-3}$) of the 
droplet and that of the ``ambient medium,'' whose
viscosity (Pa s) is called $\eta_2$. If we imagine a physical mechanism, more or less similar to a phase transition, that would separate 
CO$_2$ and N$_2$, we can estimate the velocity of a droplet of pure
supercritical CO$_2$ falling through a mixture of carbon dioxide and molecular nitrogen with typical Venusian mixing ratios, i.e., containing 
$\sim 3.5$\% of N$_2$. In this situation, using our dedicated equation of state \citep{duan_etal_1996}, under typical venusian thermodynamic conditions,
i.e., $P= 100$ bar and $T= 700$~K, we found a density of $76.1$~kg~m$^{-3}$ for pure supercritical CO$_2$, and $74.9$~kg~m$^{-3}$ if
$3.5$\% of nitrogen is added to the mixture. Concerning the viscosity, we simply took the aforementioned computed value, i.e., 
$3 \times 10^{-5}$~Pa~s (see Sect.~\ref{moldiff}). 
Then, adopting a particle diameter of $D_1 \sim 1$~mm, and
considering a value of $8.87$~m~s$^{-2}$ for the surface gravity of Venus,
we found a velocity $V_{\rm s} \simeq 7.4 \times 10^{-2}$~m~s$^{-1}$. This value enables us to estimate the timescale $\tau_{\rm CO_2,set}$ 
of pure supercritical CO$_2$ settling. Fixing, as previously, the size of the deep atmosphere to $H\sim 7 \times 10^{3}$~m, we derived 
$\tau_{\rm CO_2,set} \sim 10^{5}$~s, corresponding to $\sim 26$ hr. The timescale $\tau_{\rm CO_2,set}$ is smaller than 
$\tau_{\rm dyn} \sim 2 \times 10^8$~s
by several orders of magnitude. Even if the result depends on the droplet sizes, it is clearly seen that
the timescale characteristic of ``CO$_2$-enriched rains'' may be compatible with the timescales imposed by CO$_2$/N$_2$ dynamical mixing.
We now focus our attention on the physical mechanisms that could produce
CO$_2$-enriched droplets.
%
\begin{figure}[!t]
\begin{center}
\includegraphics[width=8 cm, angle=0]{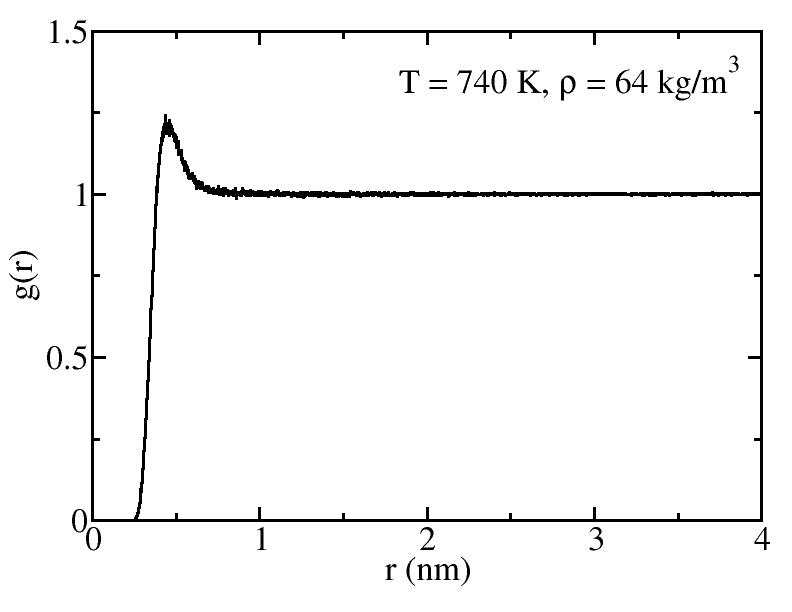}
\end{center}
\caption[]{\label{radialdistrib}Radial distribution functions between CO$_2$ center of masses at $T = 740\rm\,K$ with $\rho = 64\rm\,kg/m^3$.
           A gas-like behavior is observed: no oscillations of $g(r)$ about $1$ are apparent.}
\end{figure}
%
 Though much research is still needed in the field of supercritical fluids, stimulated by countless 
industrial applications and the issue of CO$_2$ capture and long-term storage, a substantial assemblage of literature dealing with supercritical carbon 
dioxide is available. On the experimental side, \cite{hendry_etal_2013} and \cite{espanani_etal_2016} described laboratory 
experiments involving CO$_2$--N$_2$ binary mixtures under supercritical conditions. In these experiments, the fluids are introduced in a cylindrical 
cell with an inner height of $18$~cm, the pressure is chosen between $100$ and $290$~bar, and the entire system is roughly at ambient 
temperature \citep{hendry_etal_2013}. Starting with a bulk chemical composition of $50$\% CO$_2$--$50$\% N$_2$, after a 
transitional regime of around $60$~s, the authors claim that a strong compositional gradient appears with a typical value, in mole fraction, 
around $400$\% m$^{-1}$. Although the mentioned laboratory conditions differ significantly from those found in the deep Venusian atmosphere, this result 
may suggest the existence of a fast and efficient separation process on Venus. 
 
     Unfortunately, there are several arguments questioning the reality of such a separation under the reported experimental conditions.
First, we have some doubts about the way the authors stirred the mixture inside the equilibrium cell: did they use a stirring 
device or simply rock the cell for a time until reaching equilibrium? These procedure details are not specified in \cite{hendry_etal_2013}.
Some aspects of the described experiments are clearly questionable. 
For instance, for a mixture with overall composition of $0.75$ N$_2$ and $0.25$ CO$_2$ in
mole fraction, the authors report an equilibrium persistence after $140$ hr (see their Fig. 8), indicating that the CO$_2$--N$_2$ nonhomogeneous 
supercritical fluid was not returning to a homogeneous state at $238^{\rm o}$C and $31$ MPa. In fact, to our knowledge, such a phenomenon has never 
been found in vapor-liquid equilibrium (VLE) experiments for N$_2+$alkane at high pressures and temperatures \citep{garcia-sanchez_etal_2007}.
Even more intriguing, \cite{westman_etal_2006} and \cite{macias-perez_2010}, who performed experiments on the N$_2$+CO$_2$ system, under conditions comparable to those described
by \cite{hendry_etal_2013}, observed only one single homogeneous fluid.
Finally, \cite{lebonnois_etal_2019}, who tried to reproduce \cite{hendry_etal_2013} and \cite{espanani_etal_2016}'s measurements, do not confirm 
the observations of these authors.\\
      On the theoretical side, past works have provided evidence for the possible existence of phase separations for binary systems such as the 
system CO$_2$--N$_2$; nevertheless, if they exist, they must occur at high pressure ({\it i.e.} above $8 \times 10^{5}$~bar) and high temperature (above 
$2000$ K) \citep[][]{ree_1986}. 
Recently, with modern molecular dynamics simulations, the concept of the ``Frenkel line'' has emerged for pure systems
\citep{bolmatov_etal_2013,brazhkin_etal_2013,bolmatov_etal_2014}. This line marks, within the classical ``supercritical domain,'' the boundary between 
two distinct regimes: at low-temperature, a ``rigid'' fluid following a liquid-like regime; and, at high-temperature, a non-rigid gas-like regime 
\citep{bolmatov_etal_2013}. These new findings suggest the existence of a kind of ``phase transition'' when the system crosses the abovementioned 
``Frenkel line'' \citep{bryk_etal_2017}. Because in these works the range of pressure for the ``Frenkel line'' concerning 
the system CO$_2$ was not explicitly mentioned, and given that we needed computation for CO$_2$+N$_2$, we analyzed our own MD simulations, already 
described for the evaluation of the molecular diffusion coefficient of N$_2$ molecules (see Sect.~\ref{moldiff}).
 In Fig.~\ref{radialdistrib}, we have plotted the radial distribution function $g(r)$ between the CO$_2$ center of masses for a representative numerical 
simulation at $T=740$\~K. Denoting $A$ and $B$ two CO$_2$ center of masses, $\langle\rho_B(r)\rangle$, the particle density of type $B$ particles at a 
distance $r$ around particles $A$, and $\langle\rho_B\rangle_{\rm local}$ the particle density of type $B$ particles averaged over all the spheres 
of radius $a/2$ ($a$: length of the simulation box edge) 
around particles $A$, the radial distribution function $g_{AB}(r)$ (abbreviated to $g(r)$) writes 
\begin{equation}
\begin{split}
g_{AB}(r) &= \frac{\langle \rho_B(r)\rangle}{\langle\rho_B\rangle_{\rm local}} \\
          &= \frac{1}{\langle\rho_B\rangle_{\rm local}}\frac{1}{N_A}\sum_{i=1}^{N_A}\sum_{j=1}^{N_B} \frac{\delta(r_{ij}-r)}{4\pi r^2}.
\end{split}
\end{equation}
It is a quantity aimed at evaluating the position of selected particles B with respect to a tagged particle A (CO$_2$ center of masses in both cases here), 
as well as identification of
the shell structure of solids (series of well-defined peaks) as gas-like behaviors ($g(r) = 1$).
The shape of the curve observed in Fig.~\ref{radialdistrib} 
is characteristic of gases. We did not find any
clues about the formation of CO$_2$ clusters, which could lead to the appearance of CO$_2$-enriched droplets. The simulation box view shown
by Fig.~\ref{molecularview} confirms the absence of clusters. This is in 
agreement with our discussion of laboratory results.\\
    Even though the field is relatively pristine, some studies investigating the properties of binary mixtures under supercritical conditions are 
available in the literature \citep{simeoni_etal_2010,raju_etal_2017}. Another demarcation within the supercritical domain
is introduced, the ``Widom lines,'' which may be multiple and delimit areas of the phase diagram where physical properties and chemical 
composition differ from one area to another. In Fig.~\ref{PTprofil}, we have indicated the phase diagram regions where the location of 
``Widom lines'' are expected; we can see that the Vega-2 ($P$,$T$)-profile is located well outside this region.

\section{Discussion and Conclusion}
\label{disconcl}

    Given the limited likelihood for the formation of N$_2$ gradients due to some intrinsic properties of the  N$_2$--CO$_2$ supercritical mixture,
we have to turn our attention to scenarios based on extrinsic processes. \\

    As observed by the {\it Magellan} mission, at the Venusian surface, volcanic features are ubiquitous \citep{grinspoon_2013}. 
Although space missions did not prove the existence of an active, global volcanic cycle on Venus, 
another type of phenomenon may have persisted at the surface.
This activity could have various indirect manifestations, among them chemical interactions
between the Venusian crust and the atmosphere.\\
 
 For example, nitrogen could be trapped at the surface by a geochemical process. On Earth and Mars nitrogen can be fixed via 
volcanism, lightning, or volcanic lightning \citep{segura_navarro_2005,stern_etal_2015}. The fixed nitrogen can then be trapped 
and accumulate on the surface under the form of nitrates or nitrites \citep{mancinelli_1996,stern_etal_2015}, though the stability of nitrates 
and nitrites on Venus remains unknown. Previous experimental studies have shown that CO$_2$ reacts with N$_2$ in the presence of electrical arcs. Thus,
such a mechanism could be responsible for N$_2$ depletion close to the surface if lightning is significant on Venus \citep{tartar_hoard_1930}. 
There may even be a mineral on Venus that could trap or incorporate N$_2$ into its crystal structure, a situation sometimes seen in phyllosilicates 
on Earth \citep{mancinelli_1996,papineau_etal_2005}. The abundance of nitrogen in the surface rock is necessary to determine if one of the 
discussed processes could explain the absence of N$_2$ in the near surface atmosphere.\\

  Under an alternative scenario, carbon dioxide could be released from the crust. 
As a basic assumption, we will ignore the problem of gas mixing in the atmosphere,
and we assess the geological flux $F_{\rm CO_2, min}$, of CO$_2$, that is required to get the gradient proposed by 
\cite{lebonnois_schubert_2017}. 
The overall picture is a certain amount of carbon dioxide substituting, molecule-by-molecule, the initial quantity 
of nitrogen. The latter is assumed to follow a constant profile ($x_{\rm N_2} \sim 0.035$) from the ground to $\sim 7000$~m.
According to this approach, we found $\sim 1.6 \times 10^{5}$~mol~m$^{-2}$ of N$_2$ to be replaced, leading to an 
average flux $F_{\rm CO_2, min} \sim 8 \times 10^{-4}$~mol~m$^{-2}$~s$^{-1}$, when adopting $\tau_{\rm dyn} \sim 2 \times 10^{8}$~s as
the relevant timescale. \\
  On the Earth, the main sources of natural carbon are volcanic. These sources can be classified into two types: 
direct degassing from eruptive volcanoes, and diffuse degassing from inactive volcanoes or crustal metamorphism processes regions
\citep{burton_etal_2013}. Measuring CO$_2$ emission with these geophysical structures, is rather challenging; accessibility and/or 
detectability of vents are often difficult. In addition, mixing in the atmosphere or dissolution in aquatic formations greatly complicates
the task. Nonetheless, over the past few decades, the issue of climate change has motivated advances in this field.
As an example, \cite{werner_brantley_2003} determined 
the CO$_2$ diffuse emission from the Yellowstone volcanic system: they found an average of $10^8$~mol~km$^{-2}$~yr$^{-1}$
($3 \times 10^{-6}$~mol~m$^{-2}$~s$^{-1}$). However, this value may hide significant local variations. For instance, in the 
Roaring Mountain zone \citep[see Fig. 6 in][]{werner_brantley_2003}, the authors found a median around $10^{3}$~g~m$^{-2}$~s$^{-1}$
 ($2.7 \times 10^{-4}$~mol~m$^{-2}$~s$^{-1}$), a value comparable to what is needed on Venus. Another case of diffuse source is Katla, 
a large subglacial volcano located in Iceland \citep{ilyinskaya_etal_2018}. While Yellowstone measurements have been performed with 
the accumulation chamber method, \cite{ilyinskaya_etal_2018} employed a high-precision airborne technique, together with an atmospheric
dispersion modeling. These investigations lead to a range of $12-24$~kilotons of CO$_2$ released per day. If brought to the surface of
 the glacier covering Katla, this flux corresponds to escape rates between $5.3 \times 10^{-6}$ and $1.1 \times 10^{-5}$~mol~m$^{-2}$~s$^{-1}$, 
values roughly one order of magnitude below 
what is required to sustain the Venusian atmospheric gradient. However, the total emission of CO$_2$ by Katla could be larger because 
aerial measurements may have not detected all the sources of emission, as carbon dioxide has also been found near outlet rivers by ground-based
gas sensors. Finally, an overall look at known cases of volcanic systems diffusing CO$_2$ \citep{burton_etal_2013} indicates the
estimations discussed in this paragraph should be representative of the Earth's context.\\

 Crater counting has revealed a globally youthful age for Venus' surface \citep{fassett_2016}. The $900--1000$ craters detected imply 
an average age between $0.2$ and $1$~Gyr \citep{korycansky_zahnle_2005}. In terms of geologic history, interpreting Venus' cratering and volcanic 
features is not straightforward \citep{fassett_2016}. Two alternative scenarios can be found in the literature:
(1) a catastrophic resurfacing, followed by a weak activity \citep{schaber_etal_1992,strom_etal_1994}; and (2) a surface evolution based on a steady
resurfacing, leading to a kind of equilibrium between cratering and resurfacing \citep{phillips_etal_1992,bjonnes_etal_2012,romeo_2013}.\\
   In the light of the aforementioned arguments, the chemical gradient proposed by \cite{lebonnois_schubert_2017} could be due to 
a global diffuse release of CO$_2$ from the crust.
The flux value differences between the Earth and Venus could be explained by the 
different geological evolution of these planets. 
   This argument is reinforced by the secular variability of Earth's
volcanic activity over the past $\sim 700$~million years \citep{mckenzie_etal_2016}, which suggests present-day Earth entered a regime of minimum 
CO$_2$ emission.
Remarkably, the suggested mechanism is compatible with both resurfacing scenarios. Together with 
potential thermal anomalies \citep{bondarenko_etal_2010,smrekar_etal_2010,shalygin_etal_2015}, the deep atmosphere chemical gradient could
be the mark of some remnant volcanic activity. The crustal average flux of CO$_2$ may be literally  the ``\textit{smoking gun}'' 
of this activity.\\
     The problem of the mixing of the CO$_2$, released by the crust with ambient air remains an open question. In the considered 
scenario, the turbulence plays certainly a prominent role. The turbulent mixing either in the first atmospheric layers \citep{monin_obukhov_1954}
or in the ocean \citep{burchard_2002} has itself been studied for decades. Comprehensive modelling of these processes in the context of the deep Venusian
atmosphere is well beyond the scope of this paper. Dedicated studies of the multi-species mixing, based on laboratory experiments, 
have already been initiated \citep{bellan_2017}. However, numerous works concerning mixing in turbulent jets \citep{dowling_dimotakis_1990} or 
turbulent plumes in natural contexts \citep{woods_2010} are available in the literature. If a certain amount of CO$_2$ is locally injected into 
the atmosphere with a non-negligible thrust, according to the similarity found for the concentration field of gaseous turbulent jets \citep{dowling_dimotakis_1990},
the abundances of ``fresh CO$_2$'' should firmly decrease with altitude, possibly yielding to the proposed gradient. 
  In all cases, the fate of this CO$_2$ needs further investigation because an average outgassing rate around $10^{-3}$ mol~m$^{-2}$~s$^{-1}$ would lead to
the doubling of the mass of the Venus atmosphere in less than $10,000$ Earth years \citep{lebonnois_etal_2019}, and neither atmospheric escape \citep{persson_etal_2018}
nor atmospheric chemistry \citep{krasnopolsky_2013}
seems to be able to compensate for such a flux.\\

     For the pleasure of the mind, and also because the history of science is full of surprises, we cannot exclude \textit{a priori} a 
nitrogen destruction (or carbon dioxide production?) due to some very exotic ``biological'' activity; several authors have already 
explored such a scenario \citep{morowitz_sagan_1967,sagan_1967,grinspoon_1997,schulze-makuch_etal_2004,limaye_etal_2018}. 
However, most of these works investigate hypotheses in which ``life'' develops in clouds, at relatively high altitude.\\

     In the future, spaceprobes like the pre-selected M5-mission concept \textit{EnVision} \citep{ghail_etal_2012,ghail_etal_2017}, 
will provide crucial data concerning the Venusian geological activity, for both the surface and the near-subsurface. Indeed, \textit{EnVision} 
has a proposed subsurface radar sounder (SRS) that could operate to a maximum penetration depth between $250$ and $1500$~m
under the surface of the crust \citep{ghail_etal_2017}. This instrument appears particularly relevant for the scientific question
discussed in this article.


\bibliographystyle{apj}
\def\sciam{Sci.
  Am.}\def\nature{Nature}\def\nat{Nature}\def\science{Science}\def\natastro{Nat.
  Astron.}\def\natgeo{Nat. Geosci.}\def\natcom{Nat.
  Commun.}\def\pnas{PNAS}\def\AnnderPhys{‎Ann. Phys.
  (Berl.)}\def\icarus{Icarus}\def\pss{Planet. Space Sci.}\def\planss{Planet.
  Space Sci.}\def\ssr{Space Sci. Rev.}\def\solsr{Sol. Syst.
  Res.}\def\expastro{Exp. Astron.}\def\jcis{‎J. Colloid Interface
  Sci.}\def\aap{A\&A}\def\apj{ApJ}\def\apjl{ApJL}\def\apjs{ApJS}\def\aj{AJ}\def\mnras{MNRAS}\def\araa{Annu.
  Rev. Astron. Astrophys.}\def\pasj{Publ. Astron. Soc.
  Jpn.}\def\apss{Astrophys. Space Sci.}\def\pasp{Publ. Astron. Soc.
  Pac.}\def\expastron{Exp. Astron.}\def\asr{Adv. Space
  Res.}\def\astrobiol{Astrobiology}\def\areps{Annu. Rev. Earth Planet.
  Sci.}\def\georl{Geophys. Res. Lett.}\def\jgr{J. Geophys.
  Res.}\def\gca{Geochim. Cosmochim. Ac.}\def\epsl{Earth Planet. Sci.
  Lett.}\def\plasci{Planet. Sci.}\def\ggg{Geochem. Geophys.
  Geosyst.}\def\rmg{Rev. Mineral. Geochem.}\def\tpm{Transport Porous
  Med.}\def\philtrans{Phil. Trans.}\def\faradis{Farad. Discuss.}\def\jcis{‎J.
  Colloid Interface Sci.}\def\jfm{J. Fluid Mech.}\def\physflu{Phys.
  Fluids}\def\pachem{Pure Appl. Chem.}\def\jpcA{J. Phys. Chem.
  A}\def\chemrev{Chem.
  Rev.}\def\nature{Nature}\def\nat{Nature}\def\science{Science}\def\natastro{Nat.
  Astron.}\def\natphys{Nat. Phys.}\def\natgeo{Nat. Geosci.}\def\natcom{Nat.
  Commun.}\def\scirep{Sci. Rep.}\def\science{Sci}\def\jced{J. Chem. Eng.
  Data}\def\fpe{Fluid Phase Equilibria}\def\iecr{Ind. Eng. Chem.
  Res.}\def\aichej{AIChE J.}\def\pt{Powder Technol.}\def\etfs{Exp. Therm. Fluid
  Sci.}\def\jgr{J. Geophys. Res.}\def\gca{Geochim. Cosmochim. Acta}\def\jcp{J.
  Chem. Phys.}\def\jpcl{J. Phys. Chem. Lett.}\def\jcis{‎J. Colloid Interface
  Sci.}\def\jcsft{J. Chem. Soc. Faraday Trans.}\def\jpcB{J. Phys. Chem.
  B}\def\jsf{J. Supercrit. Fluids}\def\enerp{Energy Procedia}\def\aichej{AlChE
  J.}\def\IECPDD{Ind. Eng. Chem. Process Des. Dev.}\def\pre{Phys. Rev.
  E.}\def\orggeoch{Org.
  Geochem.}\def\nature{Nature}\def\nat{Nature}\def\science{Science}\def\natastro{Nat.
  Astron.}\def\natphys{Nat. Phys.}\def\natgeo{Nat. Geosci.}\def\natcom{Nat.
  Commun.}\def\scirep{Sci. Rep.}\def\science{Sci}\def\jced{J. Chem. Eng.
  Data}\def\fpe{Fluid Phase Equilibria}\def\iecr{Ind. Eng. Chem.
  Res.}\def\aichej{AIChE J.}\def\pt{Powder Technol.}\def\etfs{Exp. Therm. Fluid
  Sci.}\def\jgr{J. Geophys. Res.}\def\gca{Geochim. Cosmochim. Acta }\def\jcp{J.
  Chem. Phys.}\def\jpcl{J. Phys. Chem. Lett.}\def\jcis{‎J. Colloid Interface
  Sci.}\def\jcsft{J. Chem. Soc. Faraday Trans.}\def\jpcB{J. Phys. Chem.
  B}\def\jsf{J. Supercrit. Fluids}\def\enerp{Energy Procedia}\def\aichej{AlChE
  J.}\def\IECPDD{Ind. Eng. Chem. Process Des. Dev.}\def\pre{Phys. Rev.
  E.}\def\prl{Phys. Rev. Lett.}\def\sciam{Sci.
  Am.}\def\nature{Nature}\def\nat{Nature}\def\science{Science}\def\natastro{Nat.
  Astron.}\def\natgeo{Nat. Geosci.}\def\natcom{Nat.
  Commun.}\def\pnas{PNAS}\def\AnnderPhys{‎Ann. Phys.
  (Berl.)}\def\icarus{Icarus}\def\pss{Planet. Space Sci.}\def\planss{Planet.
  Space Sci.}\def\ssr{Space Sci. Rev.}\def\solsr{Sol. Syst.
  Res.}\def\expastro{Exp. Astron.}\def\jcis{‎J. Colloid Interface
  Sci.}\def\aap{A\&A}\def\apj{ApJ}\def\apjl{ApJL}\def\apjs{ApJS}\def\aj{AJ}\def\mnras{MNRAS}\def\araa{Annu.
  Rev. Astron. Astrophys.}\def\pasj{Publ. Astron. Soc.
  Jpn.}\def\apss{Astrophys. Space Sci.}\def\pasp{Publ. Astron. Soc.
  Pac.}\def\expastron{Exp. Astron.}\def\asr{Adv. Space
  Res.}\def\astrobiol{Astrobiology}\def\areps{Annu. Rev. Earth Planet.
  Sci.}\def\georl{Geophys. Res. Lett.}\def\jgr{J. Geophys.
  Res.}\def\gca{Geochim. Cosmochim. Ac.}\def\epsl{Earth Planet. Sci.
  Lett.}\def\plasci{Planet. Sci.}\def\ggg{Geochem. Geophys.
  Geosyst.}\def\rmg{Rev. Mineral. Geochem.}\def\tpm{Transport Porous
  Med.}\def\philtrans{Phil. Trans.}\def\faradis{Farad. Discuss.}\def\jcis{‎J.
  Colloid Interface Sci.}\def\jfm{J. Fluid Mech.}\def\physflu{Phys.
  Fluids}\def\pachem{Pure Appl. Chem.}\def\jpcA{J. Phys. Chem.
  A}\def\chemrev{Chem.
  Rev.}\def\nature{Nature}\def\nat{Nature}\def\science{Science}\def\natastro{Nat.
  Astron.}\def\natgeo{Nat. Geosci.}\def\natcom{Nat. Commun.}\def\scirep{Sci.
  Rep.}\def\science{Sci}\def\jced{J. Chem. Eng. Data}\def\fpe{Fluid Phase
  Equilibria}\def\iecr{Ind. Eng. Chem. Res.}\def\aichej{AIChE J.}\def\pt{Powder
  Technol.}\def\etfs{Exp. Therm. Fluid Sci.}\def\jgr{J. Geophys.
  Res.}\def\gca{Geochim. Cosmochim. Acta}\def\chemgeol{Chem Geol.}\def\jcp{J.
  Chem. Phys.}\def\jcis{‎J. Colloid Interface Sci.}\def\jcsft{J. Chem. Soc.
  Faraday Trans.}\def\jpcB{J. Phys. Chem. B}\def\jsf{J. Supercrit.
  Fluids}\def\enerp{Energy Procedia}\def\aichej{AlChE J.}\def\IECPDD{Ind. Eng.
  Chem. Process Des. Dev.}\def\EF{Energy Fuels}\def\jacs{J. Am. Chem.
  Soc.}\def\nature{Nature}\def\nat{Nature}\def\science{Sci}\def\jced{J. Chem.
  Eng. Data}\def\fpe{Fluid Phase Equilibria}\def\iecr{Ind. Eng. Chem.
  Res.}\def\aichej{AIChE J.}\def\pt{Powder Technol.}\def\etfs{Exp. Therm. Fluid
  Sci.}\def\nature{Nature}\def\nat{Nature}\def\science{Science}\def\natastro{Nat.
  Astron.}\def\natphys{Nat. Phys.}\def\natgeo{Nat. Geosci.}\def\natcom{Nat.
  Commun.}\def\scirep{Sci. Rep.}\def\science{Sci}\def\jced{J. Chem. Eng.
  Data}\def\fpe{Fluid Phase Equilibria}\def\iecr{Ind. Eng. Chem.
  Res.}\def\aichej{AIChE J.}\def\pt{Powder Technol.}\def\etfs{Exp. Therm. Fluid
  Sci.}\def\jgr{J. Geophys. Res.}\def\gca{Geochim. Cosmochim. Acta}\def\jcp{J.
  Chem. Phys.}\def\jpcl{J. Phys. Chem. Lett.}\def\jcis{‎J. Colloid Interface
  Sci.}\def\jcsft{J. Chem. Soc. Faraday Trans.}\def\jpcB{J. Phys. Chem.
  B}\def\jsf{J. Supercrit. Fluids}\def\enerp{Energy Procedia}\def\aichej{AlChE
  J.}\def\IECPDD{Ind. Eng. Chem. Process Des. Dev.}\def\pre{Phys. Rev.
  E.}\def\orggeoch{Org. Geochem.}\def\sciam{Sci.
  Am.}\def\nature{Nature}\def\nat{Nature}\def\science{Science}\def\natastro{Nat.
  Astron.}\def\natgeo{Nat. Geosci.}\def\natcom{Nat.
  Commun.}\def\AnnderPhys{‎Ann. Phys.
  (Berl.)}\def\icarus{Icarus}\def\pss{Planet. Space Sci.}\def\planss{Planet.
  Space Sci.}\def\ssr{Space Sci. Rev.}\def\solsr{Sol. Syst.
  Res.}\def\expastro{Exp. Astron.}\def\jcis{‎J. Colloid Interface
  Sci.}\def\aap{A\&A}\def\apj{ApJ}\def\apjl{ApJL}\def\apjs{ApJS}\def\aj{AJ}\def\mnras{MNRAS}\def\araa{Annu.
  Rev. Astron. Astrophys.}\def\pasj{Publ. Astron. Soc.
  Jpn.}\def\apss{Astrophys. Space Sci.}\def\pasp{Publ. Astron. Soc.
  Pac.}\def\expastron{Exp. Astron.}\def\astrobiol{Astrobiology}\def\areps{Annu.
  Rev. Earth Planet. Sci.}\def\georl{Geophys. Res. Lett.}\def\jgr{J. Geophys.
  Res.}\def\gca{Geochim. Cosmochim. Ac.}\def\epsl{Earth Planet. Sci.
  Lett.}\def\plasci{Planet. Sci.}\def\ggg{Geochem. Geophys.
  Geosyst.}\def\tpm{Transport Porous Med.}\def\philtrans{Phil.
  Trans.}\def\faradis{Farad. Discuss.}\def\jcis{‎J. Colloid Interface
  Sci.}\def\jfm{J. Fluid Mech.}\def\physflu{Phys.
  Fluids}\def\pnas{PNAS}\def\pachem{Pure Appl. Chem.}\def\jpcA{J. Phys. Chem.
  A}\def\chemrev{Chem. Rev.}

\begin{acknowledgements}
We thank Dr. Christian Bouchot, Professor at the Instituto Polit\'{e}cnico Nacional (Mexico), for helpful scientific discussions.
The present research was supported by the Programme National de Plan\'{e}tologie (PNP) of CNRS-INSU co-funded by CNES, and also
partially supported by the French HPC Center ROMEO.
\end{acknowledgements}

\appendix

\section{Derivation of the Partial Molar Volume Equation}
\label{DerivMolVol}

   The partial molar volume $V_i$ (m$^3$ mol$^{-1}$) of species $i$ is given by \citep[see][ Eq.~15 page 885]{ghorayeb_firoozabadi_2000} 
\begin{equation}
    \bar{V}_i = \left(\frac{\partial \mu_i}{\partial P}\right)_{T,\{n_i\}}
\end{equation}
where $\mu_i$ is the chemical potential of species $i$, $P$ is the pressure, $T$ is the temperature and $\{n_i\}$ are the numbers of moles of 
the species $i \ne j$. The chemical potential for real gas mixtures may be written as
\begin{equation}
    \mu_i= \mu_i^0(g) + RT \mathrm{ln} \, \frac{f_i}{P_0}
\end{equation}
with $\mu_i^0(g)$ being the chemical potential of $i$ in standard state, $f_i$ the fugacity of $i$, and $P_0$ being the reference pressure.
Then the derivative is
\begin{equation}
    \left.\frac{\partial \mu_i}{\partial P}\right|_{T,\{n_i\}}= RT \frac{\partial}{\partial P}\left\{\mathrm{ln}\, f_i - \mathrm{ln} \, P_0\right\}
        = RT \, \left.\frac{\partial \mathrm{ln} \, f_i}{\partial P}\right|_{T,\{n_i\}}
\end{equation}
By definition of the fugacity coefficient $\Phi_i$ we have 
\begin{equation}
    f_i= \Phi_i P_i
\end{equation}
where $P_i$ is the partial pressure provided by
\begin{equation}
    P_i= x_i P
\end{equation}
with $x_i$ as the mole fraction of $i$ and $P$ as the total pressure. Straightforwardly,
\begin{equation}
    \left.\frac{\partial \mathrm{ln} \, f_i}{\partial P}\right|_{T,\{n_i\}} = \left.\frac{\partial \mathrm{ln} \, 
            \Phi_i}{\partial P}\right|_{T,\{n_i\}} + \frac{1}{P}
\end{equation}
leading to the desired equation
\begin{equation}
    \bar{V}_i= \frac{RT}{P} \left\{ 1 + P \, \left.\frac{\partial \mathrm{ln} \, 
            \Phi_i}{\partial P}\right|_{T,\{n_i\}} \right\}
\end{equation}      
\end{document}